# CHALLENGES AND ISSUES IN COLLABORATIVE SOFTWARE DEVELOPMENTS

**Robail Yasrab[1], Javed Ferzund[2], Saad Razzaq[3]**

1 *Dept. of Computer Sciences and Information Technology, UOS, Sargodha, Pakistan.*
2 *Dept. of Computer Sciences and Information Technology, COMSATS University, Pakistan.*
3 *Dept. of Computer Sciences and Information Technology, USTC, China.*

## ABSTRACT

Software development process has evolved with respect to the problems in developing large and complex applications. There is a paradigm shift towards collaborative development, which necessitates the need to evaluate this approach. A number of tools are used for collaborative software development (CSD) including social media and web 2.0 features. Collaborative development facilities are provided by IDE's and project hosting websites. In this paper we present a survey of collaboratively developed projects and discuss challenges and issues in CSD. We analyze various issues of communication, coordination, support, lifecycle management and discuss their effect on software quality.

*Keywords.* Collaborative Development, Project Management, Communication Tool, Open Source Software.

## 1. INTRODUCTION

Software development has turned out to be complex job that cannot be managed alone. The overall system development process is a complicated set of tasks, every one necessitating high and specific procedures and skills to be carried out. It is admitted fact that all of the levels in software development involve collaboration. Thus, the development professionals are educated and trained to become extremely technical and there is a rising feeling of lack of better handling and training on task coordination, project management, social and collaboration aspects [1].

Effective communication and collaboration is one of the important elements of the software development lifecycle. Especially, when software is developed through online social network it becomes more critical to incorporate effective communication and collaboration facilities in software development process. However, software development process through online social networks is less flexible as compared to physical software development process since there are lots of communication limitations in online working environment. Thus, due to these limitations there emerge a lot of issues and problems regarding software quality, team collaboration, task management, etc. We discuss the shortcomings and issues in the present online collaborative software development framework.

Now a day, the internet is the highly successful and extensively utilized way for sharing and accessing information. It offers us a solid ground that is used for assessing the issues and difficulties regarding the globalization of the software development industry. Presently we are having distributed software engineering and development teams all over the world. However, such type of distributed development is seen in case of large software projects. In case of collaborative and distributed development arrangement, it requires special attention. In this scenario software engineers should be able to collaborate and communicate from distant location. Moreover, for the software development businesses, internet is offering a great deal of support for the establishment of collaboration among geographically distributed development.

In collaborative development there are lots of dependencies, because collaboration development projects are globally distributed with critical parameters like distributed stakeholders, functional and technological requirements, respective information resources etc. [7].

Although distributed software development has gained fame in business organizations, it still faces some major challenges. There are some problem areas that need to be addressed. Sengupta et al. [13] highlighted some of these issues concerning knowledge acquisition and management, testing in a distributed set-up and process and metrics.

In order to deal with this kind of issues, contemporary IDEs (integrated development environments) and software project portals integrate a lot of characteristics to support coordination, collaboration and communication. These also include facilities for collaborative development like mailing lists, a version control, issue tracking and web2.0 supported functionalities. However, in offering all these functionalities and facilities in collaborative development there are a lot of problems and issues which have to be dealt with by the developers. In this paper we discuss the problems in collaborative development and in social network based development and also suggest recommendations and improvements for them. This paper will also specify the best practices those can be incorporated for the efficiency and betterment of overall development lifecycle and coordination among the collaborative system developers.

## 2. RELATED WORK

The present global technology platforms and worldwide E-Commerce depend greatly on collaboratively developed open source software. For example, programming languages such as Java, Perl, GCC, Python, Tk/TCL, operating systems such as BSD Unix, Linux, web-servers such as iPlanet/Netscape, Apache and e-mail servers such as Send mail. According to Storey et al, the collaboratively developed software systems are not considered as reliable and quality systems. Thus, the corporations, business management and decision makers usually depend upon the commercially developed systems that have been generally closed source as well as proprietary systems. However, the major reasons of less frequent practice of open source and collaboratively developed software are the lack of proper system development and various quality aspects. Moreover, the biggest problems and issues detected are about collaboration, self-organizing and communication [14].

Collaborative development demands better coordination, awareness of team activities, social awareness of the interests and opinions of all the stakeholders, awareness of the workspace along with project. Communication is also a critical component of the collaborative development, particularly when developing and implementing modern software systems at large scale that have to convene the requirements of a varied group of stakeholders and clients [1].

Normally, Software Development projects are proved to be both risky and costly endeavor. Additionally, the less effective management of the software development continues to result in missed deadlines or overall project failure. On the other hand, these problems become more complex in collaborative development environment. Kramer et al. stated that in such development environment there are social dependencies because of project's worldwide distributed nature. In addition, the collaborative development involves some critical dependence in mutual collaboration and communication [7].

In collaborative development, the role of social networks is really significant. Current social network websites are equipped with rich technology artifacts; those are having a heavy contribution to collaborative and distributed software development. Additionally, the collaborative software development always requires high degree of potential information for facilitating distributed projects through sufficient information supply [7].

Distributed collaborative system development involves a lot of project risks and quality control factors. Additionally, the on-time delivery of the collaborative software projects is considered as one of the fundamental measures of project success and quality. However, in order to get on-time delivery of the project, main requirement is to define exact project time estimation and appropriately assign the roles to project personnel [16].

As Storey et al. [15] stated that in order to improve the collaborative system development process, today's online social media is playing a significant role. In this scenario present online social media offers a variety of techniques and tools for the better software development. For instance, Web-2.0 based communication and collaboration tools and mechanisms influence software development practices. All this is happening due to paradigm shift that involves the decentralization of computer systems as well as extensive range of novel social media tools that are being accepted and customized by this new generation of developers/users.

Araujo et al. [2] assessed that sharing experiences and knowledge and combining them to carry out new actions towards a development and enhancement of systems always offer better results. Thus, offering new and better ways of communication and collaboration to software development professionals to discuss and share their experiences by setting-up communities of practice guarantees enhanced and improved performance. Majority of CSD applications are yet having a huge number of problems regarding quality and support [9].

Our work is similar to others work in identifying the challenges and issues in CSD. However, we make an extensive survey to validate the previous findings and prepare some recommendations for CSD.

## 3. SURVEY OF COLLABORATIVE SOFTWARE DEVELOPMENT RESOURCES

In order to determine overall issues regarding the collaborative development we have tried to assess several aspects of online projects and their performance. For this purpose we have selected some well known open source development websites and extracted different project features from these sources. The statistics of these features are presented in Table 1. A survey was conducted to gather fundamental features about these websites. The main websites we have used for this study are described briefly.

*SourceForge* is a web based open source software development community network. This web based platform facilitates people to develop the leading resource for open source software community. It is presently offering a lot of online software development facilities and facilitating 2.7 million open source developers to develop over 260,000 projects with more than 46 million clients [17].

*OSOR* offers facilities like technology based guidance, news, contacts, links, and work on an open source software projects repository or Collaboration Development Environment.

*Advogato* is a very old website and one of the initial trend-setters for the new ways of communication. It is a community website devoted to free open systems development. This website is also one of the initial social networking platforms that incorporated the most recent entries from every user's diary collectively into single news and project information feed that is known as recent-log.

*Eclipse* is a free, open source and modern Java applications development platform. Eclipse is also offering an IDE (Integrated Development Environment) because it offers advanced system development tools to handle workspaces, to launch, develop and debug software. Eclipse also offers a great facility of sharing software application and source code with a development team as well as other website members [19].

*Forge OW2* is another online open source collaborate development community also known as Object-Web Forge. Through such type of open source software development platform we are able to access the OW2 technical platform that comprises Subversion, CVS, bug tracking, mailing lists, task management, message boards/forums, permanent file archival, site hosting, overall web based administration and complete software backups [20].

*Gorge* is a system development website built in order to improve the communication and collaboration in open source applications development for the software community. It offers a comprehensive configured development system through versioning, a project website and system for communication among members of a software development team. It also permits the development of an open source development knowledgebase [22].

*Kenai* is a web based collaborative foster hosting website intended for free and open source systems. The Project Kenai was established by the Sun Microsystems and at the present owned by Oracle. It facilitates developers to discover each other's interest and collaboration, as well as it presents free software project hosting [23].

*Launchpad* is an online open source application that facilitates in software development and maintenance. There are lots of characteristics that Launch pad facilitate in formulating the free software also that facilitate in enhancing the overall quality of open source application development.

These characteristics comprise features similar to Trac and Bugzilla. Moreover, it offers information about the system development bugs [24].

| Website | Members (Num of online members) | Bugs (Num of bugs reported) | Download (Num of downloads) | Total blog entries (in last 12 months) | Total news updates (in last 12 months) | News Support | IM Support | Live Chat Support | RSS Feed Support | Blogs Support |
|---|---|---|---|---|---|---|---|---|---|---|
| SOURCEFORGE | 3687 | 2479 | 2,800 | 2195 | 1693 | Yes | No | No | Yes | Yes |
| OSOR | 61 | 122 | 185 | 121 | 101 | Yes | No | No | Yes | Yes |
| ADVOGATO | 9 | 44 | 59 | 59 | 37 | Yes | No | No | Yes | Yes |
| TIGRIS | 25 | 129 | 172 | 137 | 120 | Yes | No | No | Yes | Yes |
| CODEPLEX | 469 | 69 | 214 | 72 | 72 | Yes | No | No | Yes | Yes |
| ECLIPSE | 72 | 990 | 1200 | 990 | 702 | Yes | No | Yes | Yes | Yes |
| FORGE.OW2 | 47 | 113 | 136 | 1 | 86 | No | No | No | Yes | Yes |
| JAVAFORGE | 55 | 139 | 156 | 142 | 108 | Yes | Yes | No | Yes | Yes |
| GFORGE | 14 | 32 | 46 | 21 | 10 | Yes | No | No | No | Yes |
| KENAI | 35 | 64 | 80 | 56 | 0 | No | No | No | Yes | No |
| LAUNCHPAD | 72 | 610 | 709 | 510 | 0 | No | No | No | Yes | No |

**Table 1: Statistics of Online Collaborative Development Web Sites**

The reason for selecting these websites is the easy availability of data. Further these websites fulfill our objective to evaluate the collaborative development process and tools. There are only few online social network based collaborative development platforms as compared to other entertainment, games, music, commerce and communication based websites.

## 4. PROBLEM AREAS IN COLLABORATIVE ENVIRONMENT

Software development projects frequently prove to be both risky and costly. Additionally, less effective system development process and project execution cause either missed deadlines, or increase the project cost that result in overall projects failure. However, a large number of these problems come from differences among the software architecture and process model at the project planning stage to what in-fact occurs at the system development stage [1].

In this scenario, teamwork is the time complex and challenging task. The complexity and challenge of building software through teamwork can be reduced by making use of groupware to collaborate and synchronize the intricate details which are required in overall development process. At the present, businesses are more globalized which means workers can work at different locations. This globalization brings a lot of advantages and improvements to the organizations and on the other hand it also creates a lot of challenges and issues for the organizations. For instance, effective management of communication among geographically dispersed developers and staff members. In their coordination and communication success can rely on use of efficient groupware systems. Efficient and well-integrated collaborative systems promote successful coordination, communication and collaboration in an organization [3].

The social and collaborative development environment is facing some problems and issues in case of effective system development. These problems are of different nature ranging from security aspects to lack of better communication. In this section we have pointed out some of more specific issues in case of using collaborative systems for the collaborative software development.

### A. Difficulty of Evaluation

Collaborative and groupware systems are more difficult to assess than the systems utilized by individuals for the reason that they are not influenced by the personalities or backgrounds of other group members. In addition, lab circumstances cannot consistently capture complex however significant social, economic, motivational and political dynamics of collaborative development groups [2].

### B. Security Issues

In collaborative networks, personal information security is one of the main issues that occur when a hacker gains illegal access to someone's personal or private information or website's protected written language or coding. This results in the hacking or theft of such information [10].

### C. Privacy

In social and collaborative networks an attacker attacks on personal identity of someone. This type of attack is often considered as identity theft attack, where an attacker hacks someone's identity and uses it for wrong means. Spywares, worms and viruses are aimed to crawl through security protocols of social and collaborative platforms causing a constant danger for developers [4].

### D. Information Access

In collaborative development platforms the access to information and strategy of information access control varies from network to network. Some collaborative networks offer a complete access to all information and data contents while some of them offer restricted access and demand confirmation from the profile owner. Difficulties about the data and information access could be totally managed and handled by the user however illegal access could still happen particularly in case of those users who are just about starting to recognize the policy and rules of different collaborative development environments [10].

### E. Updated Notification

Each collaborative network provides a real time notification usually via email if there are any changes in the user's profile. However when these notifications are used in a negative way then they become problematic for the social network user [6].

*F. Misuse of Collaborative Networks*

Using collaborative platforms for identity theft, scams and other illegal purposes is an illegal activity. Such activities always threaten the web users and discourage the use of web. Such misuse of social platforms can lead to some drastic consequences [2].

## 5. PROBLEM AREAS IN SOCIAL NETWORKS BASED DEVELOPMENT

In collaborative software development through social network platform we can also face a lot of problems and the analysis of those problems is the basic intention of our work. This section will cover discussion on some of the main problems in open source collaborative software development through the social networks:

*A. Lack of Developers*

Even with the notable achievements of a number of open source and collaboratively developed projects, it is a harsh reality that a large number of such projects fail to take-off and thus discarded. One of the major causes behind failure of such systems is the lack of developers' interest in the collaboration [5].

*B. Poor Communication*

To make software development fruitful there is need for the better communication between project teams; though in socially developed open source projects the main difficulty is the lack of communication among development teams of project. However, the lack of communication can be due to some of the social development website features which do not allow a developer to avail the better coordination and communication facility [5].

*C. Lack of Innovation*

Another terrible reality regarding open source and online collaboratively developed is that these system are reverse-engineered and poorly written copies of existing commercial software. Novelty is hardly a mere feature in collaborative developed projects.

*D. Poor Quality*

The majority of open source and free software are unusable or poor in quality. This is because good software is developed when more than one high-quality programmers work extensively and full time collectively over a period of time in order to build, manage and improve it. Thus, in case of web based free of cost systems development on a distant social network with less effective communication and collaboration facilities it is not probable to have high quality software [15].

*E. Lack of Centralization*

System developed through a social network often lacks centralized team structure. This condition leads to less effective coordination and collaboration among development team members that results in poor quality systems. In the absence of some definite project roles, responsibility structure and proper operational management we cannot achieve high quality systems [8].

### F. Lack of Formal Mechanism

In social and collaborative development environment the overall development framework always lacks some formal system development and software engineering processes. In addition, such developments do not involve some effective methods and tool-support those are employed in traditional software development respective paradigms [8].

### G. Lack of Support

In any system application to business or corporate we often need sufficient support that permits the client and offers the freedom to improve and customize software to convene their requirements. A very small number of collaboratively developed systems are offering proper support [8].

## 6. PROPOSED SOLUTION/RECOMMENDATIONS

From the analysis and discussion of above mentioned issues we recommend the following guide lines in order to improve the quality of collaboratively developed applications. Coordination and communication is the key feature for the successful and timely completion of projects. Success rate can be increased by the use of communication facilities available on social networks, such as audio and video conferencing, instant messaging, blogs, news portals and podcasts. This communication would be available as a log along with project that can be used for future reference discussion. It can be easily analyzed and viewed for decision making.

A mechanism should be formalized to give credits to the developers of successful and quality open source application, by acknowledging their efforts and promoting that group on various social networks. This would help them in improving their profile and future endeavors. In this way we can encourage open source developers to work more devotedly and give quality software applications. Performance charts and developer profiles should be maintained along with project that helps users to check about developer's skill sets, take guidance and assess project quality.

Use of formal procedures should be encouraged in collaborative development. Success stories can be published and shared on social networks regarding use of formal procedures.

A bug tracking mechanism can provide us complete analysis from bug identification to fixing. Bug change history should be available to all developers so that any change would be seen by all collaborative developers. Further help and assistance can be obtained for locating new bugs.

Lack of centralization problem will be resolved through better and effective team coordination. This would be possible by well developed social network coordination and communication features like chatting, mobile integration with social network tools. Distributed versioning system can be used to overcome the delays in work due to multisite development. These systems allow users to work, even when they are not connected to a network. So distributed versioning system make development operation faster.

As for as market adoption and lack of support is concerned, open source developed applications like apache server, fetch mail, Linux, PostgreSQL have already captured the market due to their quality features. Promotion of this successful application should be necessary for the market adoption .Open source developers have lot of new market available for the application adoption, if they provide quality solution along with the timely support. Because the researcher community along with the new business users would always prefer the open source developed application as finances are the major focus.

A distributed project team could be successfully managed by an experienced and skilled manager. Additionally, the project manager should be capable of managing team members who are located at different locations. He must be able to deal with the people with different cultural values and background. He must be talented in resolving project team's conflicts. He should motivate his team members in order to achieve some specific goals. In this scenario, it is better to choose a project manager who has already worked in some distributed teams as a team member. It is expected from the project team members that they willingly work on distributed project. They must understand their responsibilities and put their effort in order to resolve project issues. They must take their assigned tasks seriously and work with the same loyalty as they would have done in physical working environment. They must cooperate with each other to help overcome communication barriers.

The success of a distributed project would depend on the success of task management. A project manager must understand the skills and capabilities of his team members and should assign roles and responsibilities according to their caliber.

In case of web based collaborative software development we often see lack of proper project management. In this scenario there is need for some effective development methodology like "Agile" iterative and incremental software development. Agile software development methodology offers a great deal of support for the informal communication, so it could be very effective while communicating on the web based collaborative application development platforms.

Web based instant messaging systems and chat facility should be available to make developers aware of their co-worker's present working and operating status. Video conferencing facility can be used as an alternate to face-to-face meetings.

Use of web 2.0 technologies like blogs, RSS feeds and wikis can be encouraged to improve the present communication and collaboration environment. We have found these features helpful for improving the overall software bug fixing rate. Tools like bugzilla, Bug track, MantisBT, IntelliJ IDEA can be used to make collaborative development easy and bug free.

## 7. CONCLUSION

In this paper we have discussed and analyzed some of the main problems and issues faced by the developers in the social and collaborative development environment. We investigated some of the potential issues that hinder in collaborative development of systems. We found that lack of coordination and communication facilities, poor usability, and low quality, less market adoption, lack of support and usage of formal methods are the major problem areas in collaborative software development. We have suggested some solutions and guidelines to overcome the identified problems. We recommend the use of web 2.0 tools and features for better communication and coordination. Distributed versioning system and bug tracking system should be incorporated to enhance the quality of collaboratively developed applications.